# Stabilization and frequency control of a DFB laser with a tunable optical reflector integrated in a Silicon Photonics PIC

Johannes Hauck, Matthias Schrammen, Sebastián Romero-García, Juliana Müller, Bin Shen, Jens Richter, Florian Merget and Jeremy Witzens

*Abstract* — We investigate the effect of tunable optical feedback on a commercial DFB laser edge coupled to a Silicon Photonics planar integrated circuit in which a tunable reflector has been implemented by means of a ring resonator based add-drop multiplexer. Controlled optical feedback allows for fine-tuning of the laser oscillation frequency. Under certain conditions it also allows suppression of bifurcation modes triggered by reflections occurring elsewhere on the chip. A semi-analytical model describing laser dynamics under combined optical feedback from the input facet of the edge coupler and from the tunable on-chip reflector fits the measurements. Compensation of detrimental effects from reflections induced elsewhere on a transceiver chip may allow moving isolators downstream in future communications systems, facilitating direct hybrid laser integration in Silicon Photonics chips, provided a suitable feedback signal for a control system can be identified. Moreover, the optical frequency tuning at lower feedback levels can be used to form a rapidly tunable optical oscillator as part of an optical phase locked loop, circumventing the problem of the thermal to free carrier effect crossover in the FM response of injection current controlled semiconductor laser diodes.

*Index Terms*—Distributed feedback lasers; Laser stabilization; Laser coupling; Integrated optoelectronic circuits; Hybrid photonic integration; Optical feedback; Coupled resonators

## I. INTRODUCTION

Hybrid integration of a semiconductor laser with Silicon Photonic (SiP) Integrated Circuits (PICs) allows pre-sorting of off-the-shelf laser dies with state-of-the-art performance. While progress has been made towards flip-chip laser integration [1] and alignment tolerant laser-to-chip couplers [2], [3], the issue of isolator integration has only been partially addressed in this context. Approaches based on the combination of a Mach-Zehnder interferometer with a wafer-bonded garnet typically suffer from a restricted wavelength range of operation [4]. While progress has been made in the design of high bandwidth devices [5], reflection from the first SiP chip interface may remain an issue, making more reflection tolerant lasers an important research topic for Silicon Photonics [6]. Integration of an isolator right after the semiconductor laser, with its rapidly diverging beam, has led to relatively complex packaging schemes combining a ball lens, a Faraday rotator and a laser on an optical micro-bench [7]. Significant reflections can be induced by on-chip SiP components due to the high index contrast inherent to SiP PIC technology. As a consequence, much efforts have been placed for example in the design of grating couplers with reduced on-chip back-reflections [8].

In this paper, we are investigating the effect of a tunable reflector used to compensate another unwanted back-reflection occurring elsewhere on the chip (the parasitic reflection) and shift the laser back to the desired operation regime. It is a well-known fact that the effect of back-reflections on laser dynamics depends very much on the distance between the laser and the source of the reflection [9]-[11]. The goal here is to compensate reflections occurring relatively close to the laser, i.e., at the chip interface, on the SiP chip, or at the outcoupling devices. Remote reflections occurring downstream in the fiber link need to be independently addressed with an isolator.

Another application of the tunable reflector is to shift the optical frequency of the laser by a few GHz. This can help aligning the emission wavelength to the ITU grid as an alternative to individual fine thermal control. More fundamentally, it also provides a mechanism to tune the frequency of a laser within an optical phase locked loop (OPLL) [12]. When tuning the laser frequency of a standard off-the-shelf single section semiconductor laser by modulating the injection current, an inversion in the FM response occurs at a frequency that is typically on the order of one to a few MHz. At modulation frequencies below the cross-over point, the thermal response of the laser dominates (higher current results in a reduced optical frequency). Above the cross-over, the thermal response rolls off and the free carrier effect dominates resulting in the opposite FM response. This limits the maximum operation frequency of the OPLL and prevents locking if the free running linewidth of the laser exceeds the cross-over frequency. For this reason, more complex lasers with an independent phase tuning section not subjected to thermal effects are typically implemented [13]. External control of an off-the-shelf laser by a tunable back-reflection allows

This work was supported by the European Commission under contract 619591 (Project "Broadband Integrated and Green Photonic Interconnects for High-Performance Computing and Enterprise Systems").

The authors are with the Institute of Integrated Photonics of the RWTH Aachen, Sommerfeldstr. 24, 52074 Aachen, Germany. Matthias Schrammen is now with the Institute of Communications Systems of the RWTH Aachen. Corresponding author e-mail address: jhauck@ iph.rwth-aachen.de.

circumventing this problem, as this tuning mechanism can be made high-speed (ultimately only limited by the laser and external cavity dynamics) and does not suffer from the aforementioned frequency dependence.

## II. DESCRIPTION OF PHOTONIC CIRCUIT

Figure 1A shows the schematic of a high-speed tunable back-reflector in which the power and phase of the reflected light can be independently adjusted. Light is coupled by a tunable, ring-based add-drop multiplexer (ADM) [14] to a Sagnac loop that reflects it back to the laser. The magnitude of the reflection can be set by tuning the ring resonance in and out of the free-running laser frequency. The phase is adjusted by a phase tuner interposed between the ADM and the Sagnac loop (including compensation for the phase shift introduced by the ring detuning).

Structurally similar heterogeneously integrated III-V on Silicon lasers relying on tunable ring based reflectors [15] have already been shown to provide a means to tune the emission wavelength of a laser in a Silicon based technology platform. Here, the architecture is intended to remain compatible with off-the-shelf laser diodes and relies on a comparatively weak level of resonant optical feedback compared to the aforementioned external cavity laser. While resonant feedback from external narrow linewidth resonators has been shown to provide, by itself, the means to drastically reduce the linewidth of DFB lasers [16], [17], the relatively low quality factor of the ring resonators used in this work did not result in significant improvement of the laser linewidth, as discussed in the following. Thus we are proposing to complement the relatively low quality factor of the resonators by active resonance tuning with feedback provided by an OPLL in order to achieve linewidth reduction.

Instantiating several such subsystems on a same bus waveguide would allow providing feedback to several lines of a semiconductor mode-locked laser [18]. Effect of this multiline optical feedback on the laser's free spectral range, overall spectral shape and mode-locking regime will be the subject of future investigations in view of stabilizing such lasers for their use in integrated WDM transceiver [19], [20].

Figure 1B shows the schematic of the actual PIC that was utilized to evaluate this tuning / stabilization scheme. Here, amplitude and phase are both modified by one of the micro-rings, which is thermally tuned into the laser resonance. Independent tuning of the phase was obtained by slight variations of the laser to chip distance (controlled by a piezoelectric actuator). Moreover, the thermal tuning mechanism limits the tuning speed. While this proved to be adequate to demonstrate laser frequency tuning, the demonstration of a higher speed optical oscillator for an OPLL will have to rely on high speed ring and phase tuners [21]. The reflection was induced at the end of the waveguide connected to the drop-port of the ADM by introducing a focused ion beam (FIB) cut with an estimated reflectivity of $R_{FIB} = 0.28$. A grating coupled monitor port was available to monitor the tuning of the rings.

The loaded quality factor of the tuned ring was determined to be Q=12000 on a separate break out structure and the on-resonance input- to drop-port insertion loss of the ADM to be 0.5 dB. The resonance frequency of the ring depends on the power dissipated in the thermal tuner and thus on the square of the heater current with an experimentally determined resonance frequency shift of $\Delta f_r(I) = -8.7\ GHz/mA^2$. The insertion losses of the inverse taper at the laser to chip interface and the cumulative waveguide losses between the inverse taper and the FIB cut are estimated as being respectively 2.3 dB and 0.4 dB. Thus, the maximum back-reflection that can be induced by the tunable back-reflector is estimated as -11.9 dB, close to the peak reflectivity of -11.1 dB measured by positioning a fiber instead of the laser in front of the edge coupler (which is expected to be somewhat higher because of the better matching of the lensed fiber to the mode size of the utilized edge coupler). This number ought to be compared to the reflection induced at the input facet of the SiP chip estimated to be below -14.6 dB based on the refractive index contrast between air and silica: Due to the small Silicon core of the tapered waveguide (220 nm by 200 nm) and the highly delocalized mode at the interface, the latter essentially behaves like an air/$SiO_2$ interface. Reflections into a lensed fiber from a similar edge coupler have been measured to be below -20 dB even at very small distances [2]. Moreover, from the perspective of laser dynamics the back-reflection from the tunable reflector is largely dominant if the transmission through the ADM is maximized, due to the much longer path length that plays a primary role in the determination of the effective feedback strength (see below). The cumulative waveguide length $L_C$ between the edge coupler and the FIB cut is 1.37 mm, resulting in an edge coupler to edge coupler group delay of 40 ps (group index 4.38). An additional 40 ps have to be added when the laser frequency is on resonance corresponding to the light transiting twice through the ADM.

We use a commercial strained layer multiple quantum well buried ridge stripe (SLMQW-BRS) DFB laser (1953LcV1

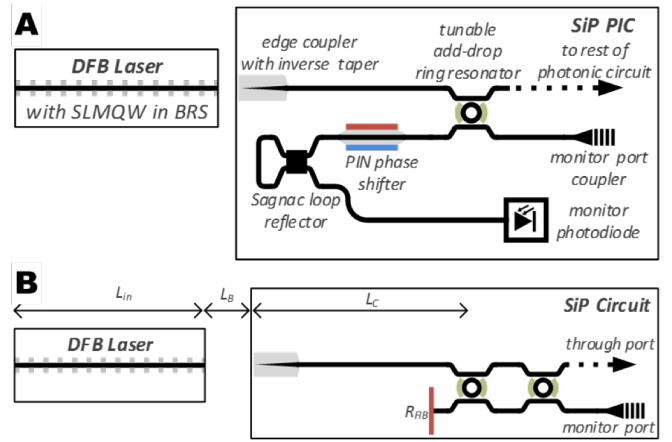

Fig. 1. (A) Schematic of the laser stabilization subsystem: a frequency selective feedback path can be rapidly phase tuned via a PIN phase shifter. (B) shows the actual chip used for the experiments. Two rings based ADMs drop light to a same waveguide in which a reflector has been implemented by means of a FIB cut. Both the ideal and the actual chip architectures feature a monitor port providing feedback for tuning the adjustable reflector.

manufactured by 3S Photonics) emitting at a wavelength of $\lambda = 1545.2$ nm. The laser's full divergence angle of 25° stems from a $1/e^2$ mode field diameter (MFD) of 4.4 μm, with some mismatch to the mode size of the inverse taper. The laser round trip time is estimated from the cavity length $L_{in} = 622$ μm to be $\sim \tau = 15$ ps. This is a rough estimate as the reduced group velocity associated to the Bragg grating was not taken into account. However, since the laser roundtrip time $\tau$, the laser facet reflectivity $R$ and the linewidth enhancement factor $\alpha$ are all partially interchangeable in the estimation of the feedback strengths (see below), this reduces to a single unknown quantity in the fits done below even though none of these three numbers are known precisely. This compound number can moreover be straightforwardly extracted from the experimentally estimated effective feedback strength (see below).

### III.  LANG-KOBAYASHI MODEL EQUATIONS

The effect of optical feedback on the laser diode can be derived from the well-established Lang-Kobayashi equations [10]. To model the experiments, two sources of reflection are considered [22]: a spectrally broadband reflection coming from the SiP chip facet (a parasitic reflection, since its phase is ill-controlled when attaching the laser to the SiP chip) and the filtered optical feedback from the FIB cut (the tunable reflection). If we assume stationary (constant amplitude and frequency) harmonic solutions that are shifted by $\omega_s$ from the solitary laser frequency $\omega_0$, we can derive the following transcendental equation (1a), with $\tau_B, \tau_F$ the (roundtrip) feedback delays for the broadband and filtered reflectors (in the second case excluding the group delay incurred in the ring, that is treated separately), and $\phi, \psi$ the corresponding feedback phases at the solitary laser frequency ($\psi$ also excluding the phase delay introduced by the ring). $\delta = 2Q(\omega_s + \omega_0 - \omega_r)/\omega_r$ is the normalized detuning of the laser frequency $\omega_s + \omega_0$ from the resonance frequency of the ring, $\omega_r$, with $Q$ the loaded Q-factor. $\tau_R = d(\arctan \delta)/d\omega_s$ is the time delay incurred by the light transiting through the ADM and is smaller than $2Q/\omega_0 = 20$ ps (and equal to the latter when the laser frequency is exactly on resonance). $C_B$ and $C_F$ are the normalized feedback strengths given by (1b) and (1c) with $R_B$, $R_F$ the (peak, on resonance) power reflectivities of the broadband (parasitic) and filtered reflectors. As defined above, $R$ is the laser facet power reflectivity and $\tau$ is the laser cavity roundtrip time.

$$\omega_s = -\frac{C_B}{\tau_B} \sin(\phi + \omega_s \tau_B + \arctan \alpha) \quad (1a)$$
$$- \frac{C_F}{\tau_F + 2\tau_R} \sin(\psi + \omega_s \tau_F + 2\arctan \delta + \arctan \alpha)$$

$$C_B = \frac{\tau_B}{\tau}(1-R)\sqrt{\frac{R_B}{R}}\sqrt{1+\alpha^2} \quad (1b)$$

$$C_F = \frac{\tau_F + 2\tau_R}{\tau}(1-R)\sqrt{\frac{R_F}{R}\frac{\sqrt{1+\alpha^2}}{1+\delta^2}} \quad (1c)$$

We can easily see that for large $C_B$ and $C_F$ multiple solutions are possible and thus multiple modes can exist. The equations are guaranteed to have only one solution for $C_B + C_F < 1$. For larger feedback strengths the other terms in the arguments of the sine functions, $\phi, \psi, \arctan \alpha$, and $\arctan \delta$, take a significant role in determining whether a bifurcation occurs in the laser frequency operation diagram and whether the laser enters multimode or unstable operation.

### IV.  COUPLING DISTANCE EXPERIMENT

We performed a first series of experiments in which the resonance of the ring was tuned to the free-running laser oscillation frequency to obtain the maximum level of feedback and in which the distance between the laser and the SiP chip was varied. At small distances, this primarily varies the feedback phases $\phi$ and $\psi$ with a half wavelength pitch (~772.6 nm) corresponding to a $2\pi$ phase shift, whereas $C_B, C_F, \tau_B, \tau_F$ are only undergoing minor changes. The rapid changes of the laser operation regime within this short range (Fig. 2A) is thus due to changes in the feedback phases. Large scale displacements of several tens of μm on the other hand add further delays to $\tau_B, \tau_F$ and increase coupling losses, the latter effectively lowering the reflectivities $R_F, R_B$. At these reduced feedback strengths, bifurcations and multi-mode operation disappear in the laser operation diagram and the dependence of the laser frequency on the feedback phase becomes smooth (Fig. 3). While this regime is less relevant to laser flip-chip attachment and laser butt-coupling, it is still shown here for completeness (it also mimics the situation of smaller laser to chip distances with reduced filtered feedback levels).

Figure 2 shows the effect of the feedback phase in a small displacement range (10 to 16 μm, wherein 10 μm corresponds to the smallest laser to chip distance we could experimentally reach and is close to the Rayleigh length of the beam emitted by the edge coupler). (A) and (B) correspond to a heterodyne measurement of the optical spectrum recorded at the through port, (C) shows the optical power simultaneously measured at the monitor port, see Fig. 1. The blue curves refer to the main mode (highest heterodyne signal RF power) whose optical frequency follows a saw tooth function with a period $\lambda/2 \approx 772.6$ nm. Within a subset of each of these periods, corresponding to a range of feedback phases, up to 8 other modes appear (other lines), while in the rest of the range the laser features stable single mode operation. However, the feedback strengths are sufficiently small for us to expect only up to three modes (for the laser close to the chip facet, $C_F$ is estimated as remaining below 3.2, as estimated from the duty cycle of single mode to multi-mode operation in Fig. 2B; $C_B$ is much smaller). An analysis of the mode frequencies shows that the modes belong to two categories. All the modes labeled with an 'a' can be generated from each other via four wave mixing inside the gain material (starting from the main, highest power modes plotted in blue and yellow that are predicted from the bifurcation diagram, see also Fig. 3). Similarly, all the modes labeled as 'b' can be generated from each other. We suspect this second family of modes might be seeded by lasing in another resonance of the DFB, the other edge of the photonic stop band

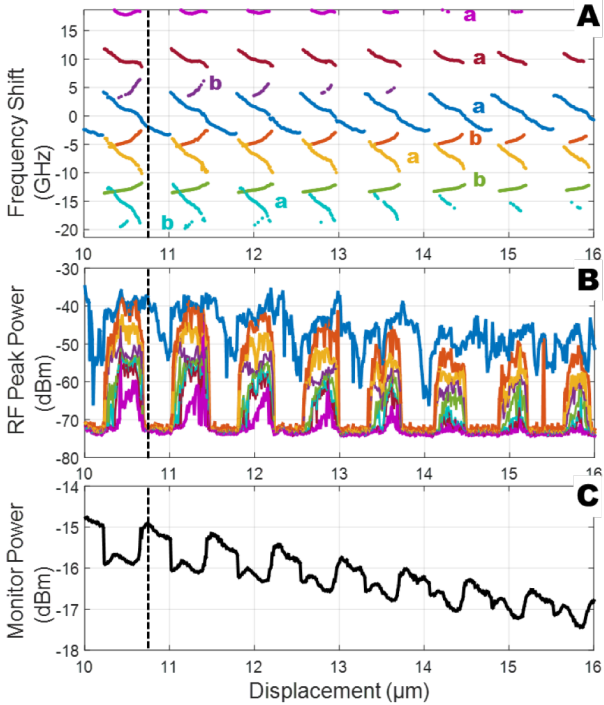

Fig. 2. Heterodyne measurement of the optical spectrum at varying coupling distances. (A) shows the mode frequencies and (B) the corresponding RF peak power recorded from the heterodyne measurement. Blue and yellow lines correspond to the main bifurcation modes, other lines also labeled with an 'a' are mixing products of these two. (C) shows the optical power measured at the monitor port, see Fig. 1. The dashed black lines show a possible operating point for a control system stabilizing the system based on the optical power recorded at the monitor port.

/ another longitudinal mode, with the opposite slope of the curve due to this second mode family having frequencies below, rather than above the reference laser (in the single sided spectrum recorded by the electrical spectrum analyzer these frequencies are folded back). The periodic drops in the peak heterodyne RF power recorded for the main line (blue) at the through port are accompanied by a broadening of the laser linewidth from ~5 to ~17 MHz as the laser operation regime gets closer to the multimode bifurcation (the integrated heterodyne power also drops, as expected close to the instability region [23], which is conducive for providing a feedback mechanism for the active reflector tuning, as discussed below).

Figure 3 shows an overlay between the measured data and the laser frequencies modeled based on equations (1a)-(1c). It is apparent that the main branch (blue and yellow curves) is modeled quite well, including the transition from a regime in which the level of reflections is high enough to trigger bifurcations to a lower feedback level for laser to facet distances above ~20 μm in which the regime stays monomode irrespectively of small displacements.

One of the outstanding issues only partially addressed here is how to best monitor the operating point of the laser and of the tunable reflector for the implementation of a control system maintaining adequate laser stabilization / parasitic reflection compensation outside of a laboratory setting (this is less of a problem in an OPLL, since it already incorporates the feedback mechanism). Obviously, an optical spectrum analyzer will not be available in a low cost device, so that alternate means will have to be found to fine-tune the tunable reflector. Exploiting effects such as the correlation between the power recorded at the monitor port and the position in the laser bifurcation diagram (Fig. 2) is a topic that needs to be further investigated. First elements are discussed here: The square wave shape of the optical power recorded at the monitor port can be straightforwardly explained: In the multimode regime, power is distributed between the different modes. Outlying modes are then filtered out by the ring resonator prior to being routed to the monitor port, so that a lower power is recorded. The optical power from the monitor port can be easily measured and is, here at least, a good indicator of the laser operating regime. In addition to being an indicator for catastrophic runoff (i.e., the laser entering multimode operation), a suitable control signal needs to also provide prior parametric information on the operating point. This can also be seen to be the case here as the square wave function is not exactly flat top, but its local maxima correspond to regions where the laser is both monomode and generates a maximized power level with a reduced linewidth.

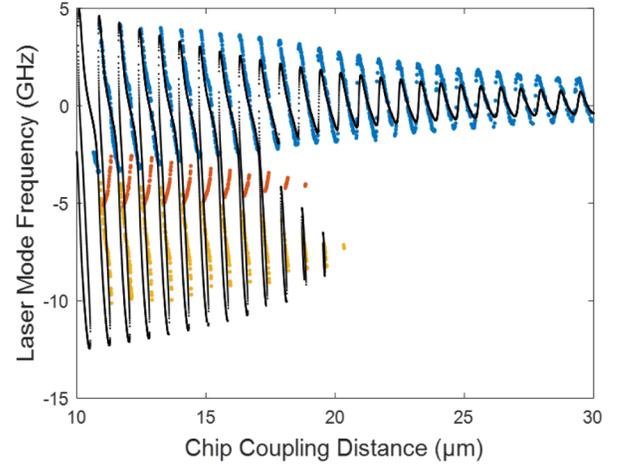

Fig. 3. Heterodyne measurement of the optical spectrum at varying coupling distances. The black dots correspond to modeling, the colored dots are experimental data following the same color coding convention as in Fig. 2.

## V. RING RESONANCE TUNING EXPERIMENTS

We conducted a second series of experiments in which the distance between the laser and the chip was fixed, but the ring resonance tuned in and out of the free-running laser frequency – the intended mode of operation with a permanently attached flip-chipped laser. In order to also independently modify the phase of the back-reflection in the absence of the PIN phase modulator depicted in Fig. 1A, we modified the laser to chip distance by small increments in a series of four experiments (nominally 200 nm to modify the phase in increments close to $\pi/2$; by fitting the data, the actual phases of the back-reflections from the tunable reflector were respectively determined to be $-1.01\pi$, $-0.45\pi$, $0.02\pi$, and $0.53\pi$). Depending on the phase of the back-reflections, smooth or more sudden transitions in the mode frequency were observed. In order to obtain high-gain but smooth operation in the feedback path of an OPLL, a high phase sensitivity can be desirable, but a regime with mode hops is obviously to be avoided. The laser to chip distance was increased to approximately ~20 µm in order to reduce the effective feedback strength $C_F$ to a regime where bifurcations do not occur (in the architecture shown in Fig. 1A this would not be necessary as phase and amplitude can be independently tuned and the feedback strength could have been independently reduced by detuning the ring).

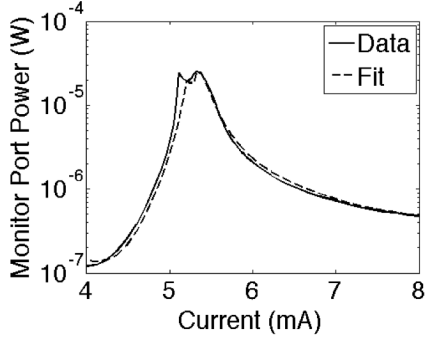

Fig. 4. Recorded and modeled power at the monitor port as a function of the current injected into the thermal tuner of the ADM.

To better understand these results, we first modeled and fitted the power recorded at the monitor port taking into account the presence of the second resonator (required to explain the asymmetricity in the transfer function of the recorded optical power as a function of the power dissipated in the heater shown in Fig. 4) as well as a small correction corresponding to nonlinear effects in the resonators (the change of the effective index inside the resonators was found to be well described by a dependency on optical power with an exponent of 1.7, between 1 and 2 as would be expected from self-heating resulting from a combination of linear and non-linear absorption [24]). While the fit for only one of the four experiments is shown exemplarily in Fig. 4, the monitor powers for all four experiments (taking into account the recorded laser frequencies) were jointly fitted, resulting in a similar quality of fit in all four cases. The parameters extracted from this fit then served as a basis to compute the phase of the back-reflection reaching the laser. Here too, common fit parameters were used to model all four data sets with the exception of a single phase corresponding to the specific displacement applied in each experiment, so that in effect all the available data sets (four bifurcation diagrams and four recorded monitor port power levels) were jointly modeled. The results are shown in Fig. 5.

In order to further give an interpretation of these results, we overlaid the modeled phase and amplitude of the reflection induced by the tunable reflector (black curves in Fig. 6 in which the tuning current of the ring is implicitly varied) with a color plot showing the predicted laser frequency as a function of this reflection. Regions in which the laser is predicted to lase with more than one line were left in white in the color plot. These regions should be avoided, as they may result in the laser not only featuring sudden mode hops during tuning, but also hysteresis and bistability. It is apparent that the laser staid in the single mode operation regime throughout all four experiments, so that the cause for the sudden transitions has to be something else. Indeed, in Fig. 6 it is further apparent that there is a sudden jump in the phase of the tunable reflection between the two flanks of the tuned resonance. This is caused by the nonlinearities inside the micro-ring and could be alleviated by reducing its Q-factor, increasing its circumference (in either case reducing the finesses and thus the optical power enhancement inside the cavity), or reducing the power levels entering the ring (for example also replacing the FIB cut by a better reflector to maintain the same feedback levels while reducing the amount of power tapped from the main signal path). Interestingly, the laser tuning curves look very different when comparing the third and fourth experiments (Figs. 5C and 5D). The phase tuning range and slope are higher in the third than in the fourth (6.7 vs. 4.65 GHz full range, 6.7 vs. 3.6 GHz in a 30 µA current tuning range), however the slope of the laser frequency vs. current is uniform on either side of the ring resonance in the fourth experiment, while the transfer function is "S shaped" in the third one. In the phase space diagrams (Fig. 6), the on-resonance back-reflection phase of the third experiment is close to zero and the tuning path almost centered on a zero back-reflection phase, while in the fourth experiment it is off-centered, leading to the aforementioned differences in the tuning characteristics (which can be seen by tracking how the black curves scan through the color plot). It is also apparent that the laser frequency differs in between the four experiments even when the resonator resonance is tuned far off the laser frequency (the heterodyne measurements shown in Fig. 5 are all referenced to the same reference laser frequency). This is due to the residual back-reflection from the chip facet, that is always present and is also taken into account in the model.

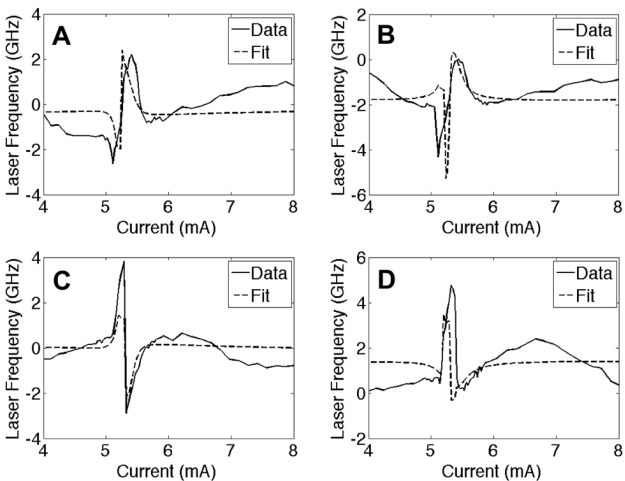

Fig. 5. Recorded and modeled laser frequency as a function of the current injected into the thermal tuner of the ADM. (A)-(D) show the results for the four experiments (corresponding to laser to chip distances modified in increments of approx. 200 nm).

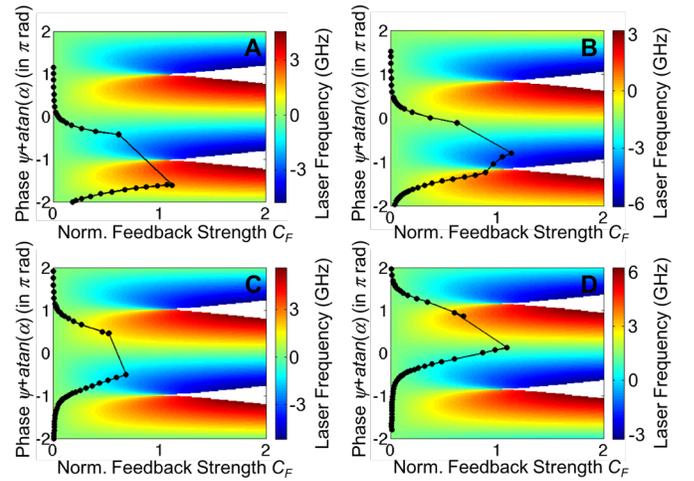

Fig. 6. The color maps show the modeled laser frequency as a function of the phase and the amplitude of the tunable reflection. The overlaid curves show the modeled characteristics of the tunable reflection (the heater current is implicitly varied). (A)-(D) show the data for the four experiments.

## VI. EVALUATION OF LASER STABILIZATION SCHEME

The first series of experiments descried earlier in the paper allowed us to validate the models describing the laser bifurcation diagram. We were also able to switch between single-mode and multi-mode operation by changing the phase of the back-reflection. However, these experiments do not yet conclusively address the compensation of parasitic reflections occurring downstream on the SiP chip by means of the tunable back-reflector. Indeed, the effect of the SiP facet back-reflection, while detuning the laser and taken into account in the modeling, was not sufficient here to trigger a bifurcation by itself, due to the small distance from the laser to the chip (rather, the bifurcation was triggered by the tunable back-reflector itself, for certain phases and high feedback strengths). We thus proceed with a numerical investigation considering parasitic (unwanted) back-reflections occurring at a larger distance. In a configuration with an isolator integrated at the output of the SiP chip, a typical source of such a parasitic back-reflection could be the interfaces and coupling devices at the output port of the chip, that would be separated from the laser by distances ranging from millimeters to a few centimeters.

Figure 7A shows the numerical evaluation of the model in which we have assumed that a -30 dB parasitic reflection, rather than being induced by the SiP chip facet in immediate proximity to the output facet of the laser diode, is induced inside the SiP chip with a path length of 6 mm separating it from the edge coupler (round trip delay time of ~172 ps). The path length between the edge coupler and the tunable reflector is also assumed to be 6 mm and the phase of the tunable reflection is set to $-\arctan\alpha$ (as the operating regime with three rather than one modes is centered on a cumulative phase of $\phi + \arctan\alpha = \pi$, a cumulative phase of zero corresponds to a stabilization against this bifurcation). The phase $\phi$ of the parasitic back-reflection is varied between 0 and $2\pi$, as the mode bifurcation should be suppressed and the variations of the laser frequency minimized for all possible parasitic phases. Beyond being a complex control problem, dynamically correcting the phase of the tunable back-reflector is not always an option even from a more fundamental perspective: Consider for example a parasitic back-reflection occurring at or close to the output of the SiP chip and a phase modulator, rapidly varying at the rate of the data, interposed between the laser and the source of the parasitic reflection. Thus, we set as a criterion that one tunable reflection phase should be adequate for all possible parasitic reflection phases.

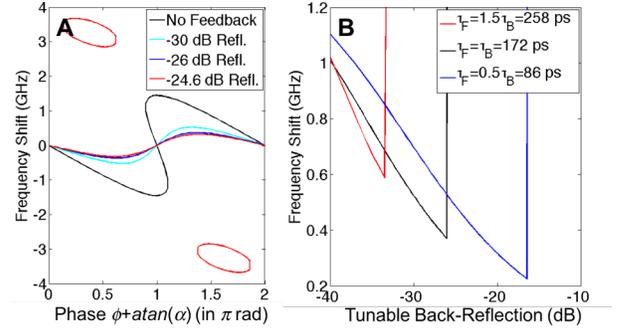

Fig. 7. (A) Laser frequency deviation as a function of the phase of the parasitic back-reflection for different tunable feedback strengths (indicated in the legend). A parasitic reflection of -30 dB is assumed. The phase of the tunable feedback is held constant at $-\arctan\alpha$ to stabilize against a 1 to 3 modes bifurcations. The time delay of both reflections is set as 172 ps corresponding to a 6 mm on-chip separation between the edge coupler and the source of the reflection. (B) Maximum laser frequency deviation (over all parasitic reflection phases) as a function of the tunable feedback strength. The black curve corresponds to $\tau_F = \tau_B = 172$ ps, the blue curve to $\tau_F = \tau_B/2 = 86$ ps and the red curve to $\tau_F = 1.5\,\tau_B = 258$ ps.

The numerical evaluation is repeated for several tunable back-reflection strengths ranging from no tunable reflection, -30 dB (same as the parasitic reflection), -26 dB, and -24.6 dB. It can be seen that as the magnitude of the tunable reflection is increased, the laser is at first increasingly stabilized. First, the three-mode bifurcation disappears. Beyond that, the variations of the laser frequency keep decreasing. However, at high feedback levels a five-mode bifurcation emerges and can occur even at a back-reflection phase of 0 (once the effective feedback strength exceeds a critical level of 4.6). This is also seen in Fig. 7A past a critical level of tunable back-reflection strength (slightly above -26 dB, that was chosen to be close to the limit after which the combined effect of the parasitic and of the tunable reflections push the laser past the 5 mode bifurcation for certain parasitic phases). This is what ultimately limits the additional level of feedback that can be artificially applied, and thus the level to which the laser can be stabilized. This also shows that the investigated scheme only works for parasitic effective feedback strengths $C_B$ significantly below 4.6, as otherwise no margin exists for introducing a tunable reflection of commensurate strength (for this laser, the maximum distance at which a -30 dB parasitic reflection can occur while maintaining its effective feedback strength below 4.6/2 is estimated to be also ~6 mm).

A remaining open question is how the performance of this scheme changes when the delay times for the parasitic and the tunable reflection are mismatched. Fig. 7B shows the worst case frequency deviations over all parasitic reflection phases as a function of the magnitude of the tunable reflection for different delay times (the delay time of the parasitic reflection is kept constant), wherein delay times equal to 50%, 100% and 150% of the delay associated to the parasitic reflection were investigated. The sharp increases in the curves correspond to the onset of the five mode bifurcation. As a general trend, it is apparent that due to the increased effective feedback strength of the tunable reflector at larger delay times, the magnitude of the laser frequency deviations can be suppressed to a higher degree at a given tunable back-reflection level. However, the onset of the five mode bifurcation also occurs at a lower reflection level, so that the net effect is a worsening of the best stabilization that can be achieved. One might then conclude that a reduction of the tunable reflector time delay would be beneficial. There is however one more aspect that needs to be taken into account: As the feedback time is decreased, higher reflection levels become necessary, which means that more light has to be tapped off from the main signal path. To place this in context, with the architecture shown in Fig. 1B, in order to obtain -20 dB tunable back-reflection, at least $1/10^{th}$ of the light has to be tapped off, as the light has to pass twice through the ADM (ignoring any other losses in the optical feedback path other than those associated to transfer through an ideal ADM). Further considering 3 dB laser to chip coupling losses and 1 dB additional waveguide losses between the laser and the Sagnac loop, the tapping ratio would have to rise to a considerable 25%. Under these conditions, the minima of the three curves in Fig. 7B respectively correspond to tapping 5%, 12% and 38% from the signal path, illustrating the limitations on reducing the length of the feedback path. This aspect should thus also be taken into account in a practical scheme (even more so if stronger parasitic back-reflections requiring higher tunable feedback strengths are being handled).

## Conclusions

In conclusion, we have investigated the applicability of a tunable reflector implemented in a SiP PIC on the stabilization and tuning of an edge coupled DFB laser. Tuning in a range of a few GHz was obtained without triggering laser mode bifurcations. This may for example find an application in the case of an OPLL operating at modulation frequencies above the thermal to free carrier cross-over in the FM response of the laser. Moreover, laser stabilization against moderate parasitic reflections occurring elsewhere on the chip at distances of a few millimeters appears realistic based on a calibrated numerical model. An outstanding challenge in this context remains the implementation of an adequate control system optimizing the settings of the tunable reflector without requiring off-chip instrumentation.